\documentstyle[aps,prb,epsf]{revtex} 
\begin{document}
\draft
\tighten
\onecolumn
\title{Strain effects on silicon donor exchange: Quantum computer
architecture considerations}
\author{Belita Koiller$^{1,2}$,
Xuedong Hu$^1$, and S. Das Sarma$^1$}
\address{$^1$Department of Physics, University of Maryland, College
Park, MD 20742-4111}
\address{$^2$ Instituto de F{\'\i}sica, Universidade Federal do Rio de 
Janeiro, 21945, Rio de Janeiro, Brazil}
\date{\today}
\maketitle
\begin{abstract}
Proposed Silicon-based quantum computer architectures have attracted 
attention because of their promise for scalability and their
potential for synergetically utilizing the available resources
associated with the existing infrastructure of the powerful Si
technology.  Quantitative understanding of and precise physical
control over donor (e.g. Phosphorus) exchange are crucial elements
in the physics underlying the proposed Si-based quantum computer
hardware.  An important potential problem in this context is that
inter-valley interference originating from the degeneracy in the Si
conduction band edge causes fast oscillations in donor exchange
coupling, which imposes significant constraints on the Si quantum
computer architecture.  In this paper we consider the effect of
external strain on Si donor exchange in the context of quantum
computer hardware.   We study donor electron exchange in
uniaxially strained Si, since strain partially lifts the valley
degeneracy in the bulk.  In particular, we focus on the effects of
donor displacements among lattice sites on the exchange coupling, 
investigating whether inter-valley interference poses less of a
problem to exchange coupling of donors in strained Si.  We show,
using the Kohn-Luttinger envelope function approach, that fast
oscillations in exchange coupling indeed disappear for donor pairs
that satisfy certain conditions for their relative positions, while
in other situations the donor exchange coupling remains oscillatory, 
with periods close to interatomic spacing.  We also comment on the
possible role of controlled external strain in the design and
fabrication of Si quantum computer architecture. 
\end{abstract} 

\section{Introduction}

\begin{figure}
\centerline{
\epsfxsize=3.6in
\epsfbox{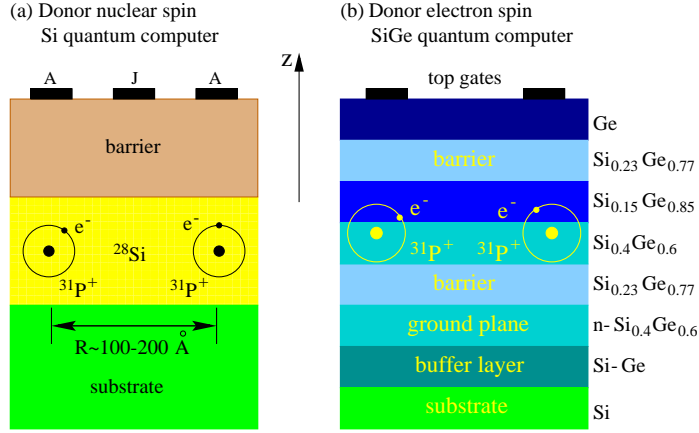}}
\vspace*{0.1in}
\protect\caption[Kane and Vrijen schemes]
{\sloppy{
(color) (a) Basic ingredients of the Silicon QC.\cite{Kane}
An array of substitutional $^{31}$P atoms in Si is
schematically represented in the figure by only two atoms.
The active elements in the QC are the the electronic and nuclear spins  
in the $^{31}$P donors. 
The nuclear spins $(I=1/2)$ are the qubits. 
The electrons mediate an effective exchange interaction between
nuclear spins and also participate in qubit initialization and
readout operations.  All operations are controlled by the A-gates,
placed above each donor, and by the J-gates, placed between
neighboring donor pairs.  A barrier layer separates the host Si
crystal from the metallic gates.  The A-gate is biased to bring the
corresponding nuclear spin in and out of resonance with an external rf
magnetic field, allowing one-qubit logic operations to be performed. 
Donor electrons mediate the nuclear spin interactions
through their exchange coupling.  Electron exchange and consequently
nuclear spin interactions are turned on or off by the
J-gate.  A uniform magnetic field $\sim 1 - 2$ Tesla is applied
along the $z$ direction.  (b) Basic ingredients of the Si/Ge
QC.\cite{Vrijen}  An array of substitutional $^{31}$P atoms in an
alloy layer of a Si/Ge composition-modulated heterostructure is
schematically represented in the figure by only two atoms.  The
qubits are the donor electron spins.  Given the difference in the
electronic $g$-factor  in Ge ($g=1.5$) and in Si ($g=2.0$),  
electron-spin resonance may be controlled by the top gate, which 
drives the corresponding donor electron into regions of different
alloy compositions.  Two-qubit exchange interactions are turned on 
through the top gates by
drawing neighboring donor electrons into a layer of small effective
mass and larger wavefunction overlap.  
}}
\label{fig:computers}
\end{figure}

Advances in quantum computing software research,\cite{QCReview}
such as the invention of factorization algorithm and quantum error
correction codes, have prompted active and extensive search for an
appropriate physical system for implementation of quantum
computation.\cite{Review}  For example, there has been considerable
recent interest in the study of donor impurities in silicon,
particularly the Si:$^{31}$P system, because of the potential of the
monovalent $^{31}$P impurities to act as fundamental units of a solid
state quantum computer (QC).  The first Si-based QC architecture,
proposed by Kane,\cite{Kane} has drawn immense interest from the
experimental community.  In Kane's proposal the nuclear spins of
phosphorus ($^{31}$P) donors are the  quantum bits (qubits), while
the donor electrons act as shuttles  between neighboring nuclear
spins.   Isotopically purified $^{28}$Si host provides a quiet
environment with very long coherence times for both donor electronic
and nuclear spins\cite{Vrijen,Kane00} since it has been known for a
long time\cite{Feher} that the electron spin coherence time in Si is
extremely long.  Two-qubit operations, which are required for a
universal quantum computer, involve precise control over 
electron-electron exchange\cite{LD} and electron-nucleus hyperfine
interactions.   A closely related alternative design was proposed
later,\cite{Vrijen} involving $^{31}$P donors in Si/Ge
heterostructures,  with the spins of the donor electrons (rather than
the nuclear spins) serving as qubits.  The basic ingredients
underlying the Si QC proposals are schematically presented and
described in Fig.~\ref{fig:computers}.  There are also other recent
QC proposals based on Si or Si/Ge host materials.\cite{Privman,Levy}
Although these are interesting
and promising proposals, particularly due to the extensive
infrastructural base available in the existing Si microelectronics
technology, there are formidable experimental challenges to be
overcome in device fabrication, coherent control, system integration,
and single spin measurement, among other important issues, before a
Si-based QC architecture could become an experimental reality even on
a laboratory scale.  

Recently, there has been significant progress in the fabrication of
donor arrays in silicon single crystal, using both a ``top-down''
approach with ion-implantation, and a ``bottom-up'' approach with MBE
growth and STM technique.\cite{Clark}  Theoretically, we have
recently shown\cite{KHD} that electron exchange in bulk
silicon has fast atomic scale spatial oscillations due to the valley
interference arising from the very
special six-fold degeneracy of the bulk Si conduction band. 
These oscillations place heavy burdens on the device fabrication and
coherent control because of the extremely high precision requirement
for placing each donor inside the Si unit cell and/or for
controlling the external gate voltages.

Several authors\cite{Vrijen,Kane00} have pointed out that Si-based 
quantum computing architectures involving epitaxial
Si/Si$_{1-x}$Ge$_{x}$ heterostructures possess some practical
advantages over the original scheme based on relaxed bulk
Si.\cite{Kane}  One of the motivations is the need for a barrier
material  separating the Si host containing $^{31}$P donors from the
conducting gates (see Fig~\ref{fig:computers}).  The
Si/barrier-material interface must be free of structural defects. 
Although high quality Si/oxide barriers can be fabricated,
Si/Si$_{1-x}$Ge$_{x}$ and Si/Si$_{1-x}$C$_{x}$ are strong candidates
for QC architectures because of the good interface quality associated
with their epitaxial nature.  These alloys have advantages over the
oxide barriers (e.g. SiO$_2$) in terms of chemical compatibility
among the group IV elements, which share the diamond lattice
structure, while still providing sufficiently high barriers for the
donor electrons to prevent leak into/from the gates.  In spite of
sharing the same lattice structure, C, Si and Ge present very large
lattice mismatches (lattice parameters 3.57 \AA, 5.43 \AA, and 5.66
\AA, respectively), leading to the presence of strained layers in
heterostructures containing commensurately grown materials.  The
highly strained nature of these epitaxial systems (arising from their
large lattice mismatch) in fact is crucial in producing high quality
defect free interface since dislocations are avoided due to the
strain build-up. 

Given our earlier finding of band-degeneracy-induced strong
oscillations in the donor exchange energy (as a function of individual
donor positioning within the Si unit cell) of bulk Si, it is natural to
ask how donor exchange behaves in strained Si heterostructure or
quantum well systems, where one must now incorporate strain effects in
the exchange calculation.  In this paper we carry out such a
calculation of strain effects on exchange energy, and emphasize an
additional feature of $^{31}$P-doped  strained Si quantum wells (QWs)
favoring such architectures over architectures based on relaxed bulk
Si for QC implementation.  Recall that two-qubit operations of a Si QC
rely on the exchange coupling between the wave functions of
neighboring donor atoms.  The six-fold degeneracy in the conduction
band edge of Si leads to oscillations in the exchange
coupling,\cite{KHD} which pose formidable experimental problems in
controlling both the fabrication (in terms of donor positioning
within the crystal) and the actual operation (in terms of surface
gate bias) of the QC.  We demonstrate below that these problems are
partially solved for impurity pairs in strained Si grown over Si/Ge
alloys due to altered lattice and band structure.  Unfortunately,
however, we find that some parts of the exchange oscillation problem
persist even in these strained systems as we discuss in this
paper.

In the following, we first discuss how to describe theoretically the
effects of strain in silicon.  We then discuss the
calculation of the donor exchange energy via an effective
Heitler-London approximation.  Finally we present our results and
discuss their implications and validity. The details of our
Heitler-London calculation are given in the Appendix.

\section{Theoretical Approach}

\subsection{Strain in Si layers}

Epitaxial growth techniques such as molecular beam epitaxy (MBE) 
allow the fabrication of lattice-mismatched heterostructures
free of misfit defect generation when the layers are
sufficiently thin (and the lattice mismatch is not extremely
large).\cite{Wang}  The mismatch is completely accommodated by
uniform lattice strain and the interatomic distances parallel to the
interfacial plane, i.e. the ``effective'' lattice constants in the
plane, remain equal to the equilibrium value of the substrate
material. The lattice constants perpendicular to the interface adjust
appropriately in order to minimize the total elastic energy in the
strained layer.  This so-called commensurate growth lowers the energy
of the interfacial atoms at the expense of increasing the elastic
(strain) energy stored in the chemical bonds in the coherently
strained layers.\cite{VandeWalle}  We refer to these layers as
uniaxially strained. Depending on whether the interatomic distances
parallel to the interface are larger or smaller than the relaxed
equilibrium value, the strain is called tensile or compressive.  The
structural variations involved in this process modify the electronic
properties of the layered materials and consequently the
substitutional impurities in the strained layers.

Here let us consider the important example of Si grown over a relaxed
Si$_{1-x}$Ge$_{x}$ substrate on a (001) interface.\cite{Kane00}  
Successful commensurate growth of such heterostructures 
has been previously reported.\cite{mobility,strain}  
The average alloy lattice parameter, which follows 
Vegard's law very closely,\cite{windl} defines the Si layer lattice
parameter parallel to the interface:  a$_\parallel$ = $(1-x)$ a$_{\rm
Si}$ + $x$ a$_{\rm Ge}$.  The lattice parameter perpendicular to the
(001) interface is also modified in the strained Si layer with
respect to its equilibrium value.  According to macroscopic
elasticity theory,  a$_\bot$ = a$_{\rm Si} + 2(c_{12}/c_{11})$
(a$_{\rm Si}$ - a$_\parallel$).  Here $c_{12}$ = 16.577 dyn/cm$^2 $
and $c_{11}$ = 6.393 dyn/cm$^2 $ are elastic constants with values
given for relaxed bulk Si.\cite{madelung}  The elasticity theory
predictions for the crystal lattice structures in Si/Ge
heterostructures were confirmed by {\it ab initio}  calculations by
Van de Walle and Martin.\cite{VandeWalle}  For an alloy with
$x=0.2$, the above relations yield a$_\parallel$ = 5.474 \AA~ and
a$_\bot$ = 5.396 \AA, i.e., distortions of less than 1\% ($+0.8\%$
and $-0.6\%$) with respect to the equilibrium value of a$_{\rm Si}$ =
5.43 \AA.

Uniaxially strained Si has lower symmetry than relaxed Si, and thus a
different reciprocal lattice (body centered tetragonal instead of
body centered cubic) and Brillouin zone (BZ).
Herring and Vogt\cite{HerringVogt} have shown that the 
small tetragonal distortions cause changes in the conduction band
minima of Si that can be quantitatively described
by shifts in the energies of the local minima of the six valleys,
while the reciprocal-space positions and shapes of the
constant-energy surfaces remain unchanged to first order in strain. 
The energy shifts are given in terms of the amount of 
distortion and of the {\it uniaxial} and {\it dilation} deformation
potentials,  $\Xi_u$ and $\Xi_d$, respectively.  For Si grown over
relaxed Si$_{1-x}$Ge$_{x}$ (001) substrate, $\Xi_d$ leads to uniform
shifts of the six valley minima.  Therefore, as discussed below, 
only $\Xi_u$ is relevant for our study of $^{31}$P donor wave functions.

\subsection{Effective Hamiltonian for a donor electron in strained
Si}

The conduction band edge of relaxed bulk Si consists of six degenerate 
minima located along the $\langle 100 \rangle$ directions, at about
85\% away from the BZ center, towards the zone boundary at X points.
Within the Kohn-Luttinger envelope function approach,\cite{Kohn}
the ground state for a donor electron in Si is also six-fold
degenerate, and the corresponding states are labeled by the
reciprocal-space points ${\bf k_\mu}$ at which the valley minima 
occur.  For definiteness, we follow the sequence: $\mu = 1,2, \ldots
, 6 \leftrightarrow z,-z,y,-y,x,-x$,  with the envelope functions
given by (e.g. for $\mu = \pm z$)\cite{Kohn}  
\begin{equation}
F_{\pm z} ({\bf r}) = \frac{1}{\sqrt{\pi a^2 b}} \ 
e^{-[(x^2+y^2)/a^2 + z^2/b^2]^{1/2}} \,.
\label{eq:envelope}
\end{equation}
The effective Bohr radii for Si from a 
variational calculation are $a = 25.09 $ \AA~ and $b= 14.43$
\AA.\cite{KHD}  

A substitutional impurity breaks the translational
symmetry of the host crystal, leading to intervalley scattering
effects known as the valley-orbit
interactions,\cite{Baldereschi,Pantelides} which split the
unperturbed six-fold degenerate donor electron ground state into a
singlet ($A_1$), a triplet ($T_1$) and a doublet ($E$), with the
ground state being a non-degenerate state of $A_1$ symmetry.  We
follow the perturbative treatment of Baldereschi,\cite{Baldereschi}
which starts from the six degenerate ground-state functions
calculated within the one-valley Kohn-Luttinger variational approach.
The low energy donor electron wave functions can be expanded in the
basis spanned by the six valleys   
\begin{equation}
\psi ({\bf r}) = \sum_{\mu = 1}^6 \alpha_\mu F_{\mu} ({\bf r})
\phi_\mu(\bf r)\,,
\label{eq:sim}
\end{equation}
where $\phi_\mu({\bf r}) = u_\mu({\bf r})e^{i {\bf k}_{\mu}\cdot
{\bf r}}$ are the pertinent Bloch wave functions and $\sum_{\mu =
1}^6 |\alpha_\mu|^2=1$.  Each $\alpha_\mu$ coefficient is in general
referred to as $\mu$-{\it valley population}, while a set of 6
coefficients defines the so-called {\em valley composition} of a 
donor electron state.  

To determine the donor electron wave function, we include as
perturbations two types of intervalley coupling, for valleys on
perpendicular symmetry directions (e.g. $x$, $z$) and for those on
opposite symmetry directions (e.g. $z$, $-z$).  We represent these
couplings by $-\Delta_C$ and $-\Delta_C (1+\delta)$, respectively. 
The experimental values\cite{Grimmeiss} for the relative splittings
among the $A_1$, $T_1$ and $E$ levels for $^{31}$P donors in Si are
obtained as $\Delta_C =2.16$ meV and $\delta = -0.3$.  Taking
valley-orbit scattering into consideration through these
parameters leads to a binding energy of 40 meV for the $A_1$ ground
state, in quite a good agreement with the experimental
value of 45.5 meV.\cite{Grimmeiss}  More accurate estimates of the
binding energy were obtained through non-perturbative variational
treatments,\cite{Pantelides} which however rely on a spherical-band 
approximation and adopt s-like hydrogenic trial envelope functions. 
Since our goal here is to obtain a realistic description of the ground
state wave function for the donor electron rather than the precise
values of the binding energy, we employ the perturbative approach
with anisotropic envelope functions.  For our considerations of Si QC
hardware architecture, which focus on electron exchange
due to wavefunction overlap, the perturbative envelope function
approximation that we employ is quite adequate.

Uniaxial strain along [001] induces different valley shifts 
for the two local minima along the $z$ axis ($\mu = 1,2$) compared to 
the other four along $x$ and $y$ axis ($\mu = 3,4,5,6$).  Within the
subspace spanned by the electron wave functions at the six conduction
band minima $\{F_{\mu} ({\bf r}) \, \phi_\mu({\bf r})\}_{\mu = 1,6}$, 
the donor ground and lower excited states may be conveniently
obtained as the eigenvectors of the effective perturbation
Hamiltonian in the form of a $6\times 6$ matrix which, for $^{31}$P
donors in a Si QW uniaxially strained along the $z$ axis, is written
as\cite{Wilson} 
\begin{eqnarray}
H = H_{vo} + H_{\rm strain} + H_{\rm z} 
%
= -\Delta_C 
\pmatrix{0&1+\delta&1&1&1&1\cr
	1+\delta&0&1&1&1&1\cr
	1&1&0&0+\delta&1&1\cr
      1&1&1+\delta&0&1&1\cr
      1&1&1&1&0&1+\delta\cr
      1&1&1&1&1+\delta&0} 
\nonumber
\end{eqnarray}
\begin{equation}
+\Delta_C 
\pmatrix{2\chi&0&0&0&0&0\cr
	0&2\chi&0&0&0&0\cr
	0&0&-\chi&0&0&0\cr
      0&0&0&-\chi&0&0\cr
      0&0&0&0&-\chi&0\cr
      0&0&0&0&0&-\chi}
+
\pmatrix{-\Delta&\Omega&0&\cdots\cr
	 \Omega^*&-\Delta&0\cr
	0&0&0\cr
      \vdots&~&~&\ddots\cr
      ~\cr
      ~}~.
\label{eq:Hvo}
\end{equation}
Note that the order of the rows and columns of the perturbation 
Hamiltonian $H$ follows the sequence in $\mu$ given above
Eq.~(\ref{eq:envelope}).  Furthermore, since our concern here is
the donor wave functions, any shift in $H$ proportional to the
identity has been neglected, so that the eigenvalues of
$H$ give the correct relative splittings among the eigenstates, but
not the binding energies.  The first term of the perturbation
Hamiltonian, $H_{vo}$, gives the valley-orbit scattering due to the
presence of $^{31}$P donors in unstrained Si
discussed above.\cite{Baldereschi}  The second term, $H_{\rm
strain}$, introduces the relative energy shifts due to uniaxial
strain along [001] in terms of a dimensionless scalar {\em valley
strain} parameter $\chi$ defined in terms of strain parameters and
the valley-orbit scattering matrix element\cite{Wilson}  
\begin{equation}
\chi = {\Xi_u e_T \over 3\Delta_C} \,.
\label{eq:strain}
\end{equation}
Here the uniaxial strain parameter $\Xi_u$\cite{HerringVogt}
is approximately 8.6 eV for the Si conduction band
edge,\cite{Balslev,Laude} while $e_T = e_{\rm zz}-e_{\rm xx}$ is
the difference between the diagonal components of the strain tensor
for uniaxial strain along $z$: $e_{\rm zz} = {{\rm a}_\bot/{\rm
a}_{\rm Si}} - 1$ and $e_{\rm xx} = {{\rm a}_\parallel/{\rm{a}_{\rm
Si}}} - 1$, so that the valley strain parameter $\chi$ is linearly
related to the Si$_{1-x}$Ge$_x$  alloy composition $x$ as  
\begin{equation}
\chi = \frac{\Xi_u}{3\Delta_C} \, \frac{{\rm a}_{\rm Si} - 
{\rm a}_{\rm Ge}}{{\rm a}_{\rm Si}} \left( \frac{2c_{12}}{c_{11}} + 1
\right) x 
\label{eq:chi}
\end{equation}
Since $\Xi_u$ is three orders of magnitude larger than $\Delta_C$ 
(with $\Xi_u \sim 10$ eV and $\Delta_C \sim 2$ meV), relatively 
small changes in $x$ may lead to important shifts in the energies 
of the valleys.  For example, for Si grown over a relaxed
Si$_{1-x}$Ge$_{x}$ alloy with $x=0.2$ on a (001) interface, lattice
distortions (smaller than 1\%) lead to a valley strain parameter
$\chi \sim -20$.  For Si$_{1-x}$Ge$_{x}$ alloys of arbitrary
composition, we estimate $\chi(x) = -95 x$. Negative values of $\chi$
indicate tensile strain $(a_\parallel > a_{\rm Si})$, favoring the
$z$ valleys energetically.  Commensurate growth of Si over
Si$_{1-x}$C$_{x}$ or SiC, which is more difficult experimentally due
to the extreme lattice mismatch between the materials, would lead to
compressive strain $(a_\parallel < a_{\rm Si})$, so that $\chi >0$. 
The lattice mismatch between C (diamond) and Si is 34\% (3.57 \AA~
versus 5.43 \AA), as compared to 4.2\% for Ge with respect to Si
(5.66 \AA~ versus 5.43 \AA).  When we replace a$_{\rm Ge}$ by a$_{\rm
C}$ in Eq.(\ref{eq:chi}),  the estimated valley strain parameter for
Si  grown over relaxed Si$_{1-x}$C$_{x}$ would be $\chi \sim 800 x$.
This is actually a lower bound for $\chi$, since 
this alloy's average lattice parameter 
is somewhat smaller than predicted by Vegard's law.\cite{windl} 

A symmetry argument based on the differentiated {\em fit} of the 
six envelope functions to the Si host geometry and boundary
conditions may be used to include the effect of the
confinement potentials due to the alloy regions in
Si$_{1-x}$Ge$_x$/Si/Si$_{1-x}$Ge$_x$ heterostructures, or due to an
interface with any potential barrier (e.g. Si/SiO$_2$) within a few
effective Bohr radii from the impurity.  Assuming that all Si/barrier
interfaces are perpendicular to the growth direction $z$, the
components with $F_{\pm z}$ envelopes are favored energetically,
regardless of the value or sign of $\chi$, since the smaller 
effective Bohr radius $b$ along $z$ guarantees a less 
significant penetration of the wave function into the barrier
regions as compared to the other envelopes with the larger Bohr
radius $a$ along $z$.  This effect is phenomenologically included in
the third term $H_{\rm z}$ of the donor electron Hamiltonian $H$
through an energy shift $\Delta$.  The parameter $\Delta$ is always
positive, and its value depends on the barrier height and on the
impurity position with respect to the interface.\cite{Bastard}  We
estimate it here to be up to $\sim 10$ meV.

In addition to valley energy shifts, the presence of an interface
perpendicular to $z$ also leads to coupling between the
$\pm z$ valleys.  Surface-induced intervalley scattering\cite{Sham}
is introduced phenomenologically in  $H_{\rm z}$ through the parameter
$\Omega$.  This coupling is in general complex, and leads to energy
shifts smaller than 1 meV in realistic
situations.\cite{Sham,Koester}

\subsection{Single donor electron states and energy spectrum}

In the absence of strain and other perturbations, the single donor
electron Hamiltonian consists only of the valley-orbit
coupling, i.e. $H=H_{vo}$.  If $\delta=0$, so that all the
valley-orbit scattering matrix elements are the same, the
eigenstates of $H_{vo}$ are a singlet with energy $-5\Delta_C$ and a
quintuplet with energy $\Delta_C$.  When $\delta \neq 0$, the
quintuplet splits into a triplet and a doublet, so that a single
donor electron has the state sequence of a singlet, a triplet, and a
doublet, which have energies $-(5+\delta) \Delta_C$, $(1+\delta)
\Delta_C$, and $(1-\delta) \Delta_C$, respectively.  For Si, where
$\delta \approx -0.3$ and $\Delta_C \approx 2.16$ meV,  the singlet
is at $-4.7 \Delta_C$, the triplet at $0.7 \Delta_C$, and the doublet
at $1.3 \Delta_C$.  The first excited state  here is a triplet that is
$5.4\Delta_C \sim 11.7$ meV above the ground singlet state.  The
valley compositions of these states show that the ground singlet has
$A_1$ symmetry, corresponding to a symmetric superposition of all six
valleys: $\frac{1}{\sqrt{6}} (1,1,1,1,1,1)$, while the triplet and
doublet have $T_1$ and $E$ symmetries as expected.\cite{Kohn}

When the Si lattice is uniaxially strained along the $z$ axis, $H =
H_{vo}  + H_{\rm strain}$.  Now the degeneracy in the triplet and
doublet  states are also lifted.  For $\chi < 0$, the ground state
energy is  $\Delta_C \left[-(2+\delta) +\chi/2 -(3/2) \sqrt{\chi^2 +
4\chi/3 +4} \, \right]$: It is an admixture of the $A_1$ state with
one of the components of $T_1$, with valley composition
$(\alpha_1,\alpha_1,\alpha_3,\alpha_3,\alpha_3,\alpha_3)$ and $\alpha_3 =
\alpha_1/\left[ \sqrt{\chi^2 + 4\chi/3 +4} - (\chi+2/3) \right]$. 
Notice that the ground state approaches $(1,1,0,0,0,0)/\sqrt{2}$ as 
$\chi \rightarrow -\infty$.  In this limit, the first
excited state is one of the original triplet states (in the absence
of strain), with energy $\Delta_C (1 + \delta + 2\chi)$ and valley
composition $(1,-1,0,0,0,0)/\sqrt{2}$.  The energy splitting
between these lowest states approaches $2(1+\delta) \Delta_C
\sim 1.4 \Delta_C \sim 3.02$ meV.  
For $\chi > 0$, the  lowest energy states are  
admixture of $A_1$ with the other two
components of $T_1$:  The (nondegenerate) ground state has energy
$\Delta_C [-(2+\delta) -\chi/2 - (3/2) \sqrt{\chi^2 +4\chi/3 +4}]
\sim \Delta_C (-3-\delta-\chi)$, while the first excited state
(actually a doublet)  has energy $\Delta_C (1+\delta-\chi)$, with a
splitting from the ground state approaching $(4+2\delta)\Delta_C
\sim 3.4 \Delta_C \sim 7.34$ meV as $\chi \to +\infty$.  
Hydrostatic (dilation) strain causes a rigid energy 
shift in the six valleys, keeping the splittings and valley
compositions of all eigenstates unchanged. 
Since our focus here is to understand wavefunction overlaps, we do not
consider effects of hydrostatic strain. 

\begin{figure}
\centerline{
\epsfxsize=3.6in
\epsfbox{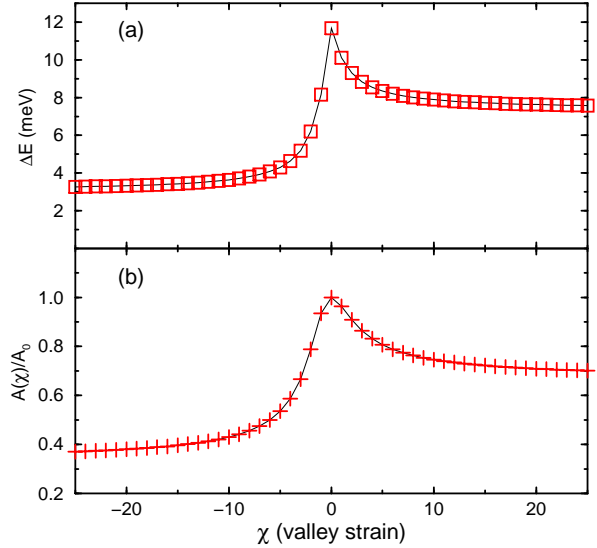}}
\vspace*{0.1in}
\protect\caption[Single donor state characteristics]
{\sloppy{
(a) Energy splitting between the ground and first excited state
of a single electron bound to a single $^{31}$P donor in strained Si
host as a function of the valley strain $\chi$.  For Si commensurately
grown on Si$_{0.8}$Ge$_{0.2}$, with $\chi \approx -20$, this energy
splitting is about 3.3 meV, down from $\sim 12$ meV in relaxed bulk
Si. For $\chi > 0$, the first excited state is a doublet.  At large
$\chi$, this energy splitting is about 7.5 meV.  (b)Hyperfine
coupling of the ground single donor electron state, normalized to the
unstrained host value, as a function of the valley strain $\chi$. 
At $\chi =-20$, the ground state hyperfine coupling is about $38\%$
of the ground state in relaxed bulk Si.
}}
\label{fig:SD}
\end{figure}

In Fig.~\ref{fig:SD}, panel (a), we plot the energy splitting between
the first excited state and the ground state of a single donor
electron as a function of the valley strain $\chi$.  
This quantity is relevant in estimating time scales which determine
the adiabatic condition in time-dependent processes driven by the gate
potential variation.  As we have shown above, the energy splitting
becomes smaller when the bulk ground and excited states mix to form 
strained ground and excited states.  The variation of these energies 
can be seen in Fig.~1 of Ref.~\onlinecite{Wilson}.  It is also
clear from that figure why the curve we plot here has a cusp at $\chi
= 0$: A level crossing occurs at $\chi=0$ for the first excited state.

The strength of the hyperfine coupling between the donor electron
and nuclear spins is also important for the Si QC,  since it is
invoked in both single qubit and two-qubit operations.\cite{Kane} 
In terms of the donor electron state composition, it can be expressed
as  
\begin{equation}
A(\chi)/A_0 = \frac{1}{6} \left| \sum_\mu \alpha_\mu \right|^2 \leq 1
\,,
\end{equation}
where $A_0$ is the value of $A(\chi=0)$.
This is a differentiable function of $\chi$, so its peak at $\chi=0$
is smooth, as is shown in panel (b) of Fig.~\ref{fig:SD}. 
Furthermore, for $\chi \rightarrow -\infty$, the ground state
composition approaches $1/\sqrt{2}(1,1,0,0,0,0)$, so that
$A(-\infty)/A_0 = 1/3$; while for $\chi \rightarrow \infty$, the
ground state approaches $1/2(0,0,1,1, 1,1)$, and $A(\infty)/A_0 =
2/3$.  The curve in Fig.~\ref{fig:SD}(b) clearly approaches these two
limits.  Results summarized in Fig.~\ref{fig:SD} indicate that donors
in uniaxially strained Si will have ``inferior''single-electron
properties compared to the relaxed material for the purpose of QC
operations, requiring longer times to perform an adiabatic evolution
of the electron state and providing a reduced hyperfine coupling.

When the $z$-direction confinement is added, it introduces very small
energy shifts and additional off-diagonal coupling between $z$ and
$-z$ directions. In the presence of $H_{vo}$ and/or $H_{\rm
strain}$, the effects of $H_z$ are negligible for all relevant
properties discussed here.  Thus we do not include this term in our
calculation below.

We emphasize the following general properties of the spectrum of 
$H$ in Eq.~(\ref{eq:Hvo}) when $\Delta_C > 0$:
(i) the ground state is always non-degenerate, and 
(ii) the ground state valley populations in opposite 
symmetry directions, e.g. $\mu =1, \nu=2$, always satisfy 
$|\alpha_\mu|^2 = |\alpha_\nu|^2$.

\subsection{Heitler-London approximation}

In this paper we use Heitler-London (HL) approximation to calculate
the inter-donor exchange coupling as the energy difference between
the two-electron singlet and triplet states.\cite{Slater}  
Considering a pair of donors at ${\bf R}_A=0$ and ${\bf R}_B={\bf R}$
and assuming $R = |{\bf R}| \gg a,b$ (the effective Bohr radii), the
energies of the singlet and triplet states are 
\begin{equation}
E_{\stackrel{s}{t}} = 2 E_0 + {H_0\pm H_1 \over 1\pm S^2} \,,
\label{eq:hl}
\end{equation}
where $E_0$ is the single impurity electronic energy, while $H_0$,
$H_1$, and $S$ are integrals involving two electronic wave functions
of the form (\ref{eq:sim}) centered at the origin and ${\bf R}$,
and are given in the Appendix.
Contrary to the case of the H$_2$ molecule, where these integrals are
isotropic and smoothly decaying functions of the relative distance
$R$, in the case of Si they are extremely anisotropic and
sensitive functions of ${\bf R}$.\cite{KHD}  The HL expression for
the exchange splitting in Si is
\begin{eqnarray} 
J({\bf R}) & = & E_t - E_s \nonumber \\
& = &  
\sum_{\mu, \nu} \Big[\sum_{{\bf K},{\bf K'}} |c^\mu_{\bf K}|^2 
|c^\nu_{\bf K'}|^2 e^{i({\bf K}
-{\bf K'})\cdot {\bf R}}\Big ]|\alpha_\mu|^2
|\alpha_\nu|^2 {\cal J}_{\mu \nu} ({\bf R}) \cos ({\bf k}_{\mu}-{\bf
k}_{\nu})\cdot {\bf R}\,,
\label{eq:exchl}
\end{eqnarray}
which is derived in the Appendix, where the explicit expression for
${\cal J}_{\mu \nu} ({\bf R})$ is also given.
The second summation (the part within the square bracket) in
(\ref{eq:exchl}) refers to the  reciprocal lattice expansion of the
periodic part of the Bloch function, $u_\mu({\bf r}) = \sum_{\bf K}
c^\mu_{\bf K} e^{i {\bf K}\cdot {\bf r}}$, and is identically unity
when ${\bf R}$ is an fcc crystal lattice vector.  

A prominent feature of the above expression for the exchange
coupling $J({\bf R})$ is the presence of the fast oscillatory terms
$\cos ({\bf k}_{\mu}-{\bf k}_{\nu})\cdot {\bf R}$, which are periodic 
with wavelengths of the order of the atomic spacing ($\sim
5$\AA~ in Si).  The coefficients ${\cal J}_{\mu \nu} ({\bf R})$ are
slowly varying functions of ${\bf R}$ as they are integrals
containing the envelope functions $F_{\mu} ({\bf R})$, whose
characteristic decay length is the effective Bohr radius of the order
of 20 \AA.

\section{Results}

\begin{figure}
\centerline{
\epsfxsize=3.6in
\epsfbox{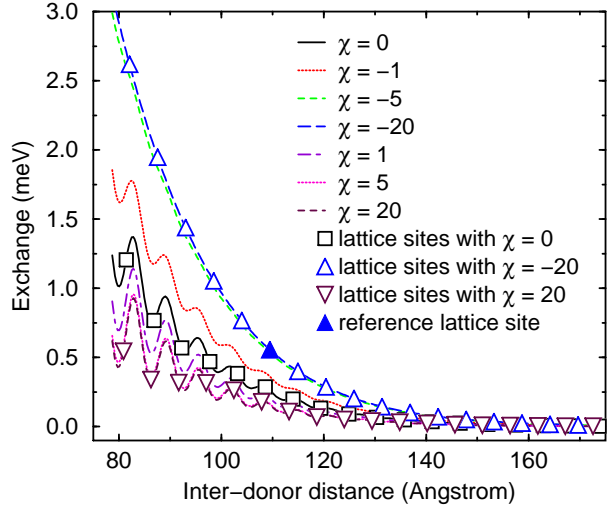}}
\vspace*{0.1in}
\protect\caption[Exchange along $x$ direction]
{\sloppy{
Calculated donor electron exchange in Si as a function of inter-donor 
distance and valley strain $\chi$, with the two donors located along
the [100] axis (with the same $y$ and $z$ coordinates).  The symbols
represent exchange values when both donors are on substitutional
lattice sites, while the curves give results for continuously varied
inter-donor distance.  The filled symbol in particular refers to the
configuration in which one donor is at the origin while the second
donor is 20 lattice spacings away along $x$ axis.  This configuration
will be used as a starting reference point for the next two figures.
Notice that for systems with different values of $\chi$, the symbols
representing lattice sites are slightly displaced horizontally 
relative to each other.  The reason is self-evident, as the strain is
produced by growing Si commensurately on substrates ranging from
Si/Ge ($\chi<0$) to Si/C ($\chi>0$) alloys, with varying alloy
compositions that lead to different lattice spacings.    
}}
\label{fig:Exch-x}
\end{figure}

\begin{figure}
\centerline{
\epsfxsize=3.6in
\epsfbox{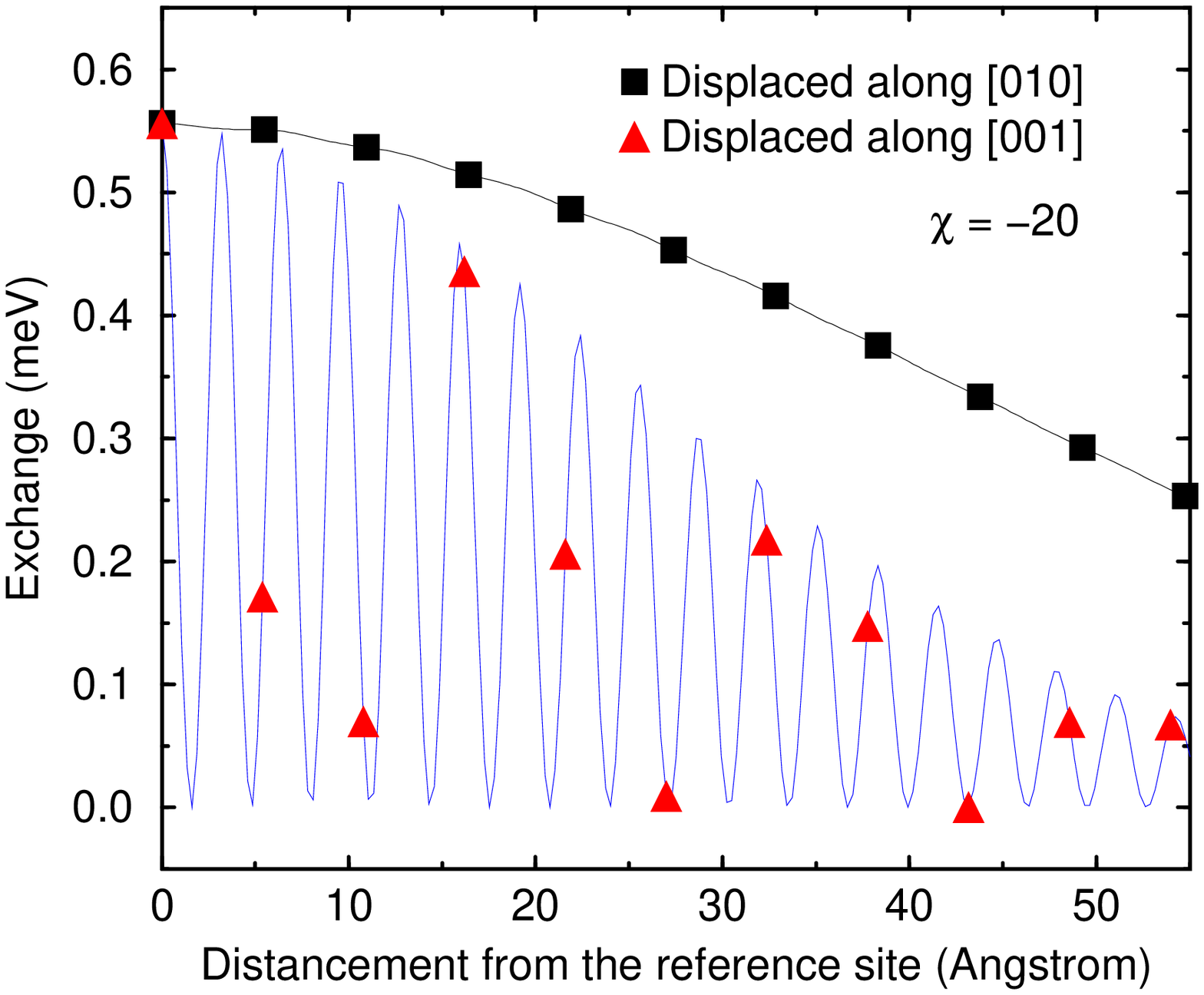}}
\vspace*{0.1in}
\protect\caption[Exchange along $y$ and $z$ directions]
{\sloppy{
Calculated exchange coupling as a function of displacements
of one of the donors relative to the reference configuration in Si
lattice uniaxially strained along $z$ axis with $\chi = -20$.  The
origin represents the situation when the two donors are in the
reference configuration defined in Fig.~\ref{fig:Exch-x}.  The two
curves in the figure refer to the cases when one of the donors is
displaced from its reference site along the $y$ and $z$ axis,
respectively.  The symbols represent the substitutional lattice
sites along the two axis.  
}}
\label{fig:Exch-yz}
\end{figure}

\begin{figure}
\centerline{
\epsfxsize=3.6in
\epsfbox{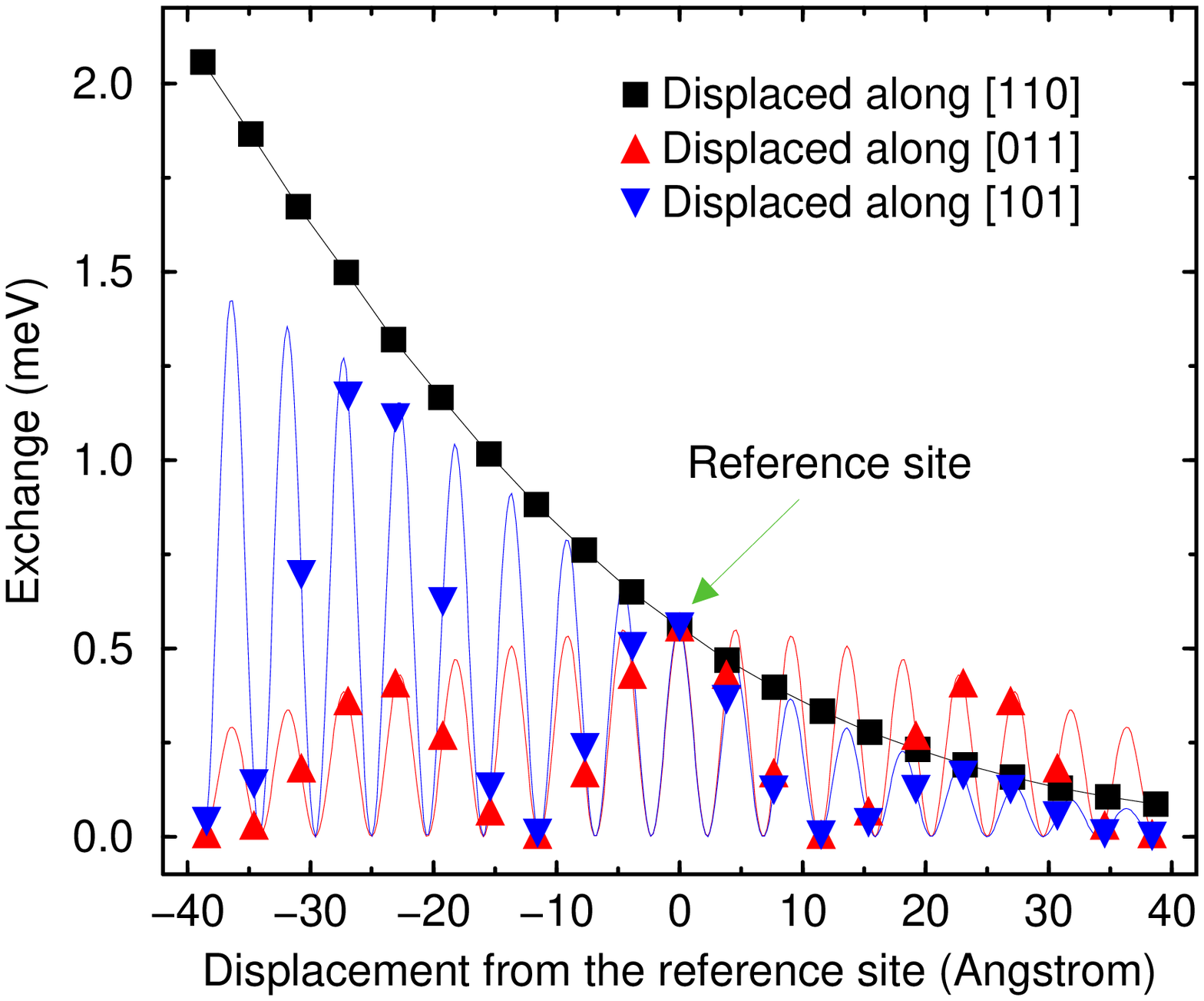}}
\vspace*{0.1in}
\protect\caption[Exchange along diagonal directions]
{\sloppy{
Calculated exchange coupling as a function of displacements
of one of the donors relative to the reference configuration in Si
lattice uniaxially strained along $z$ axis with $\chi = -20$.  The
origin represents the situation when the two donors are in the
reference configuration defined in Fig.~\ref{fig:Exch-x}.  The three
curves in the figure refer to the cases when one of the donors is
displaced from its reference site along the [110], [011], and [101]
axis, respectively.  The symbols represent the substitutional lattice
sites along the three axis.  The curve for the [011] axis is symmetric
around the reference site because the corresponding displacement
along the [011] axis causes symmetric variations in the inter-donor
distance $|{\bf R}|$. 
}}
\label{fig:Exch-Diag}
\end{figure}

\begin{figure}
\centerline{
\epsfxsize=3.6in
\epsfbox{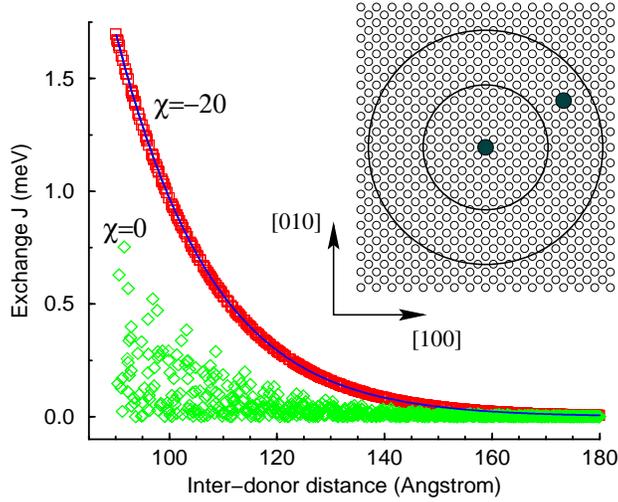}}
\vspace*{0.1in}
\protect\caption[Exchange in-plane]
{\sloppy{
Calculated exchange coupling for in-plane displacements
of the donors in the $x$-$y$ plane for 
relaxed Si ($\chi=0$, diamond symbols) and strained Si
(uniaxial strain along the $z$ direction.  $\chi = -20$, square
symbols).  The inset, in which the filled dots represent the donors, 
describes schematically the positions considered for the displaced 
donor, which consist of all possible lattice sites between two
concentric circles of radii 90 \AA~ and 180 \AA~ with the
other donor positioned at the center of the circles.  The data points
correspond to the exchange calculated at all
relative positions considered.  The solid line is $J({\bf R})$ for
${\bf R}$ along the [100] direction for $\chi = -20$.
}}
\label{fig:Exch-plane}
\end{figure}

We have performed HL calculations of the donor electron exchange 
for different configurations of a pair of donors.  
Results (Figs.~\ref{fig:Exch-x}-\ref{fig:Exch-plane}) presented here
are obtained by taking the summation over reciprocal lattice vectors
to be unity: $\sum_{{\bf K},{\bf K'}} |c^\mu_{\bf K}|^2 
|c^\nu_{\bf K'}|^2 e^{i({\bf K}-{\bf K'}) \cdot {\bf R}}=1$, i.e. the
term within the square bracket in Eq.~(\ref{eq:exchl}) is taken to be
unity which is equivalent to calculating $J({\bf R})$ only for those
values of ${\bf R}$ which correspond precisely to ${\bf R}$ being an
fcc lattice vector.  The continuous lines shown in the figures
correspond to the free-electron  limit $c^\mu_{\bf K} \approx
\delta_{{\bf K},0}$,  which is a first order approximation to the
lowest conduction band  in homopolar semiconductors.\cite{Walter}  We
investigate the exchange coupling in uniaxially strained Si as we
vary the inter-donor distance, the inter-donor direction relative to
the strain axis (defined to be the $z$ axis), and the sign and the
magnitude of strain itself.  Results presented in
Figs.~\ref{fig:Exch-x} to \ref{fig:Exch-plane} give a comprehensive
account of the general trends and effects to be expected, providing
new qualitative and quantitative insight into the problem of donor
electron exchange coupling in multivalley semiconductors.   

In Fig.~\ref{fig:Exch-x}  we plot the exchange coupling $J$ as a
function of the inter-donor distance ${\bf R}$ and valley strain
$\chi$ for a pair of donors along the  [100] axis.  Notice that as
$\chi$ increases from negative to positive values, the exchange
coupling becomes increasingly oscillatory.  Such behavior can be
easily understood from the expression of $J({\bf R})$ in
Eq.~(\ref{eq:exchl}) and from the donor ground state valley
composition.  For $\chi \ll 0$ with strain along the $z$ axis, the
dominant components of the ground state are  
\begin{equation}
\psi({\bf r}) \sim \alpha_1 F_{z} e^{i {\bf k}_z \cdot {\bf r}}
+\alpha_2 F_{-z} e^{-i {\bf k}_z \cdot {\bf r}}~,
\label{eq:wavef}
\end{equation}
where $\alpha_1 = \alpha_2 \sim 1/\sqrt{2}$, so that the largest
oscillatory contribution to $J({\bf R})$ comes from terms with
factors $\cos (k_z \pm k_{-z})R_z$ with $k_{-z} = -k_z$.  The only
effect of varying the $x$-component of ${\bf R}$ is to vary the
coefficients ${\cal J}_{\mu \nu}$, which  are slowly decaying
functions of $|{\bf R}|$.  As $\chi$ increases toward zero, the
contribution from the four components $\alpha_3, \ldots, \alpha_6$
grows, and fast oscillations are superimposed on the overall decaying
envelope of $J(R_x)$.  For $\chi \gg 0$, these four components
dominate the ground state valley composition, and the oscillations are
enhanced with respect to relaxed bulk Si. 

From the perspective of predicting and controlling the donor pair
exchange coupling, results in Fig.~\ref{fig:Exch-x} indicate that
the optimal degree of strain is in the range $\chi
\lesssim -5$, since this leads to a simple hydrogenic-type behavior
for the particular type of relative positions (along [100] axis) of a
pair of donors, as was presumed in the original Si and Si/Ge QC
proposals.\cite{Kane,Vrijen}  Moreover, the value of $J$ is
considerably enhanced (by about $100\%$) with respect to values in
the relaxed bulk Si, a feature with considerable practical advantages
in terms of increasing the speed of gate operations in the QC.

To study cases when the two donors are not positioned along the [100]
axis, we first choose a reference configuration, with a donor pair
along [100] direction.  We choose to work with uniaxially strained Si
with $\chi = -20$ from now on as it corresponds to the realistic
situation of Si grown over relaxed Si$_{0.8}$Ge$_{0.2}$.  We
arbitrarily pick the relative position $R_x \sim 110$\AA~ marked by
the solid triangle in Fig.~\ref{fig:Exch-x}, with respect to which
various displacements are considered for one of the two donors. 
In Fig.~\ref{fig:Exch-yz} we plot the exchange coupling $J$ as one of
the donors is displaced from the reference site along either the $y$
([010]) or the $z$ ([001]) axis.  The upper curve, corresponding to
displacements along the $y$ axis, reflects essentially the same
qualitative behavior as in Fig.~\ref{fig:Exch-x} for $\chi =
-20$.  However, when the relative displacement ${\bf R}$ acquires
{\it any} nonzero $R_z$ component, exchange becomes rapidly
oscillatory, as indicated by the triangles in the lower curve.  The
same trends are seen in Fig.~\ref{fig:Exch-Diag}, where
displacements along the [110], [011] and [101] directions are
considered.  Among those, only  displacements along the [110] diagonal
direction lead to a monotonous variation of $J$ with distance
without any oscillations.  The other two types of displacement
involve a nonzero $R_z$ component in the relative  position $\bf R$,
leading to fast oscillations in the exchange energy.  

The overall effect of uniaxial strain in the exchange coupling of
a donor pair in the $x$-$y$  plane is summarized in
Fig.~\ref{fig:Exch-plane}.  There we consider all possible in-plane
substitutional donor pairs with inter-donor distances in the range
of 90 \AA~ $<R<$ 180 \AA.  For impurity pairs in uniaxially strained
Si $(\chi = -20)$, the exchange is a single-valued function of $R$,
decaying smoothly with distance, while for unstrained Si $(\chi = 0)$
the exchange is significantly reduced with respect to the strained
results, and is sensitively dependent on the relative position {\bf
R}. 

Notice that in all the results presented here the
oscillatory behavior of $J$ is almost always accompanied by partial
cancellations, meaning that in the oscillatory regime, the values of
$J$ are always smaller than in the corresponding non-oscillatory
situations.  This is clearly evident in Fig.~\ref{fig:Exch-x},
comparing curves with increasing values of $\chi$; or in
Figs.~\ref{fig:Exch-yz} and \ref{fig:Exch-Diag}, comparing curves
with zero and nonzero $R_z$ components for the same inter-donor
distance $|\bf R|$.  This result can be mathematically  confirmed
from  expression (\ref{eq:exchl}),  where the $\cos({\bf
k}_{\mu}-{\bf k}_{\nu})\cdot{\bf R}\,$ factors are responsible both
for oscillations and for partial cancellations [${\cal J}_{\mu
\nu} ({\bf R})>0$ for arbitrary $\mu$ and $\nu$].  It is
somewhat different (and perhaps counterintuitive) 
from an ordinary two-wave interference.  One aspect of our results
presented in the figures (Figs.~\ref{fig:Exch-x}-\ref{fig:Exch-plane})
that needs to be emphasized is that our exact HL calculations are done
only for discrete values of the relative displacement or inter-donor
distance, and the smooth curves are guides to the eye.  As mentioned
in the beginning of this section, the smooth continuous curves in our
results correspond to an effective free electron approximation.

\section{Summary and Discussions}
\subsection{Approximations}

We now briefly review the main approximations involved in the
theoretical approach adopted in this paper, and discuss various
possible improvements.  In our perturbation theory calculation, the
only effect of strain is to change the valley populations of
the donor electron states through the diagonal elements of the
six-valley Hamiltonian $H$.  We do not consider strain-dependent
changes in the effective masses, the inter-valley scattering matrix
elements $\Delta_c$ and $\delta$, or the BZ locations for the
conduction band minima.  Since the lattice parameter
changes are quite small---less than 1\% as we mentioned in the
beginning of Sect.~II, we believe that our approach is well-suited
for the modestly strained systems considered here.  For highly
strained systems (such as Si grown on Si$_{1-x}$C$_x$ with high
concentration of carbon), the separation of $H$ into $H_{vo}$ and
$H_{\rm strain}$ may not be justifiable, and all matrix elements
should be consistently assessed to determine how they depend on
strain.

In our envelope function approach, we have adopted 
the variational form and parameters obtained from the 
one-valley Kohn-Luttinger treatment. 
Calculations incorporating the valley-orbit interactions
and strain in the original Hamiltonian, and treating the effective 
Bohr radii $a$ and $b$ and the valley compositions as independent 
variational parameters (imposing symmetry and normalization
constrains), would have led to a better quantitative 
description of the ground state binding energy and possibly 
the wave function.  Such calculations for donors in Si have been
performed to determine the binding energies using only
spherically symmetric trial wave
functions,\cite{Pantelides,Oliveira} which is not adequate for the
present study.  Our interest in this paper is not in obtaining
extremely accurate donor binding energies, but in calculating
reasonably good wavefunctions so as to obtain reliable donor exchange
energies.

Atomistic {\it ab initio} studies based on density-functional schemes 
can overcome the limitations of the envelope-function/effective-mass
theory and have been employed in the literature to study effects of
strain\cite{VandeWalle} and alloy disorder.\cite{windl} 
The difficulty in describing a single shallow impurity originates 
from the underlying supercell approach with periodic boundary
conditions,  which leads to spurious effects due to interactions
between the impurity  and its periodic replicas. 
Extrapolation schemes towards an infinite supercell have been
proposed for the single impurity binding energy,\cite{menchero}
although periodic boundary conditions cause additional difficulties
for a reliable description of the  asymptotic behavior of the wave
functions. 

The HL approximation adopted in this paper is the simplest
approach to calculate exchange coupling in a variety of situations,
such as atomic lattices,\cite{Herring} random donor
arrays,\cite{Andres} and double quantum dots.\cite{BLD,SHD} 
Improvement can be made through various ways.  For instance, it is
quite common to use the valence bond theory (also called the
molecular orbital theory) or the configuration interaction (often
combined with a self-consistent mean field calculation) approach to
obtain more accurate spectra (and thus exchange interaction when
specifically pursued) for molecules or quantum
dots.\cite{Slater,BLD,SHD,Maksym,HD}  For donor arrays in Si
(relaxed or strained), such calculations will be complicated by the
Si band structure.  However, as we have shown in the current study,
as long as the effective Bohr radius is much larger than the lattice
spacing, one can separate the fast oscillating (due to the Bloch
functions) parts of integrals for the matrix elements from the slowly
varying envelope function parts, thus simplifying the calculation. 
Indeed, there should be no significant theoretical difficulty for a
more complete molecular orbital calculation for donors in Si.  In the
current study, our focus is on the qualitative behavior of the
exchange coupling between two donors, therefore we have not attempted
to improve upon the HL approximation.  This can be easily done
following earlier calculations in GaAs quantum dots\cite{HD} if the
experimental development warrants such improvements of the theory.  In
addition, the proposed Si QC would operate  under a uniform magnetic
field of 1 to 2 Tesla  applied along the $z$ axis (see
Fig.~\ref{fig:computers}). Previous studies\cite{HD} indicate that
uniform magnetic  fields of up to 2 Tesla have no qualitative effect
in the exchange coupling of  electrons in double quantum dot bound
states.  For a pair of donors, a reduction in $J$ due to the
shrinking of the envelopes perpendicular to the magnetic field
direction and magnetic phase interference should be expected, but
this should not lead to any qualitatively different effect. 

As we have mentioned before, in our calculation we have approximated
the Bloch functions by the corresponding plane wave functions, which
is in essence a free electron approximation for the
conduction band of Si.\cite{Walter}  The approximation here amounts
to keeping only the $c_0$ term (and let $c_0=1$) in the reciprocal
lattice expansion of the periodic part of the Bloch functions.  The
plane wave approximation actually obtains the exact results for
lattice points on the same fcc lattice, where $e^{i({\bf K}-{\bf
K}')\cdot {\bf R}}=1$ so that the sums over reciprocal lattice vectors
${\bf K}$ are normalized to 1.  However, for an arbitrary $\bf R$,
our calculation should mostly give an {\em upper bound} to an exact HL
calculation.  If the $c_{\bf K}$ coefficients have a broad
distribution, the sums would tend to be much smaller since they
involve the fourth power of $c_{\bf K}$ as is shown in
Eq.~(\ref{eq:exchl}).  Thus the continuous lines in our figures
(Fig.~\ref{fig:Exch-x} to Fig.~\ref{fig:HL-A}) should be taken more
as guides to the eye than as the actual values of exchange, which can
be much smaller.  As mentioned before, we carry out our exact
Heitler-London exchange coupling calculations only for discrete
values of inter-donor separation (triangles, squares, etc. in the
figures) whereas the continuous curves represent the free electron HL
approximation. Further calculations are underway to go beyond the
free electron approximation to address this issue, and those results
will be presented elsewhere.

\subsection{Valley degeneracy effects}

Our results for Si uniaxially strained along the $z$ axis indicate
that,  for $\chi \lesssim -5$, inaccuracies in the positioning of
substitutional $^{31}$P donor pairs within the $x$-$y$
plane produce only small effects in the electron exchange
coupling.  Naturally, the next question is whether it is possible
to avoid the oscillatory behavior due to
lattice-parameter-scale inaccuracies in the $R_z$ component of the
relative position between a pair of donors.  As noted in Sec.~III,
these oscillations in $J({\bf R})$ are due to the specific form of the
ground state wave function (\ref{eq:wavef}), which dictates that the
contribution from the two valley minima are equally
important.  This result is in turn a consequence of the degeneracy of
the ${\bf k}_z$ and $-{\bf k}_z$ valleys in bulk Si, whose energies
are lowered equally with respect to the four perpendicular valleys by
uniaxial tensile strain.  The same effect occurs in the
two-dimensional electron gas (2DEG) in (001) Si/SiO$_2$ inversion
layers\cite{Pudalov} and Si/Si$_{1-x}$Ge$_x$ quantum
wells\cite{Koester} where, due to the confining potential at the
interface, the lowest electronic subbands originate from the bulk Si
${\bf k}_z$ and $-{\bf k}_z$ valleys alone.  It has been shown
theoretically\cite{Sham,Grosso} and
experimentally\cite{Koester,Pudalov} that this valley degeneracy may
be lifted in heterostructure-type configurations, leading to
experimental {\em valley splittings} of the order of 0.1 meV.  Such
splittings refer to the unperturbed doubly-degenerate ground states, 
and the resulting singlet ground state is not one of the  ${\bf k}_z$
and $-{\bf k}_z$ valleys, but an equally weighted admixture of these
states.\cite{Sham}  In the present case, $H_{vo}$ alone leads to a
nondegenerate ground state,  which is 12 meV below the first excited
state.  Under the $H_{\rm strain}$ perturbation, the ground state
remains nondegenerate,  with an energy separation to the first
excited state reduced, but still larger than 3 meV.  Further
perturbations, e.g. $H_z$ that produces the valley splitting 
observed in the 2DEG, have negligible quantitative  impact in any of
the results obtained here. Moreover, perturbations of the form $H_z$
always lead to an equally weighted admixture of the ${\bf k}_z$ and
$-{\bf k}_z$ valleys, thus donor electron valley composition always
has contributions from these two valleys with the same weight, i.e.
$|\alpha_1| = |\alpha_2|$ in (\ref{eq:wavef}).

We note that for Ge, where four conduction band minima occur at 
the inequivalent $L$ points in the BZ, $\langle 111 \rangle$ tensile 
strain leads to a nondegenerate band minimum.  
Since the lattice parameter of Ge is larger than that of Si and C, 
this possibility within group-IV semiconductors remains restricted
to Ge-rich strained alloys grown over relaxed Ge,\cite{Vrijen}
with the additional complication of bringing alloy disorder effects
into the non-degenerate ground state layer where the donors are
placed.  We do not consider Ge in this paper since there is no
particular experimental effort directed toward the design of a
Ge-based QC.

\subsection{Estimates for static electric field effects} 

Static electric fields generated by the top gates can be used to
slightly modify the donor electron wavefunction in its shape and
weight distribution, so that such fields can be used to overcome
lattice-parameter-scale inaccuracies in the $R_z$ component of the
relative position of a pair of substitutional impurities.
A crude estimate of the uniform static electric field effects 
in the donor bound states may be inferred from results for the
hydrogen atom.\cite{schiff}  To the first order in the field,
the H ground state wave function in the presence of a uniform 
electric field $\bbox{\cal E}$ along $z=r\cos\theta$ is given by 
\begin{equation}
\psi_{\rm H}({\bf r}) = \frac{1}{\sqrt{\pi a_0^3}} \; e^{-r/a_0}
\left[ 1 -\frac{\cal E}{e} \left(a_0 r + \frac{r^2}{2} \right)
\cos\theta \right]\,, 
\label{eq:field}
\end{equation} 
where $a_0 = 0.529$\AA~ is the Bohr radius, and $e$ is the electron
charge.  The expectation value of the electron position is thus
shifted by the field from  $\langle {\bf r} \rangle = 0$ to  
$\langle {\bf r} \rangle_{\rm H} = -\frac{9}{2} a_0^3 {\vec {\cal
E}}/{e}$, which corresponds to an average displacement of $4.63\times
10^{-6}$\AA~ for ${\cal E} = 10$ kV/cm.  This negligible shift is
greatly enhanced for donor bound states in Si.  Assuming a
spherically averaged envelope with an effective Bohr radius $a^* =
(a^2b)^{1/3}$, we get $\langle {\bf r} \rangle_{\rm Si} =  \epsilon
(a^*/a_0)^3 \langle {\bf r} \rangle_{\rm H} \cong  7.4 \times 10^5
\langle {\bf r} \rangle_{\rm H}$, where  $\epsilon = 12.1$  it the Si
dielectric constant.  An average position shift of one monolayer (1ML
= a$_{\rm Si}$/2 = 2.7 \AA~ in Si) thus corresponds to ${\cal E} =
8$ kV/cm. In the current context, it is plausible to expect that a
uniform electric field of tens of kV/cm may compensate for atomic
misplacements of a few monolayers along the field direction.

It is interesting to note that, according to Eq.(\ref{eq:field}), 
the applied field would have no first-order effect on the 
electron-nucleus hyperfine coupling, 
which is proportional to the electronic density at the nuclear 
site, $A \sim |\psi(0)|^2$.
Since the A-gates are primarily meant to reduce this coupling by   
pulling the electron wavefunction away from the donor towards the 
barrier, fields much stronger than those quoted above 
are required to perform such operations. 
In the strong field limit, donor ionization starts to take place.
The field dependence of the dissociation rate for atomic H
is\cite{oppenheimer} 
\begin{equation}
\frac{\Delta N }{T} \propto  {\cal E}^{1/4} \exp \left(-
\frac{2e}{3{\cal E} a_0^2} \right), 
\label{eq:rate}
\end{equation} 
which means that dissociation becomes appreciable for 
field values in the order of ${\cal E}_{\rm H} =  e/a_0^2 = 5\times
10^6$ kV/cm.  Under the spherically averaged envelope assumptions
above, we estimate that donor ionization in Si would 
require fields of the order of ${\cal E}_{\rm Si} =
[(a_0/a^*)^2/\epsilon] {\cal E}_{\rm H} \cong 5.3 \times 10^{-5}
{\cal E}_{\rm H} \sim 300$ kV/cm, which is one order of
magnitude larger than the fields required to compensate
for atomic displacements of a few ML.

\subsection{Fabrication-related aspects}

We have shown in this paper that strained Si grown commensurately 
over relaxed Si$_{1-x}$Ge$_x$ (001) alloys presents 
clear theoretical advantages over relaxed Si as 
a host material for $^{31}$P donor impurities in  
Si-based QC architectures. 
Since the same is not true for strained Si grown on Si$_{1-x}$C$_x$
substrates, here we focus on growth of strained Si over
relaxed Si$_{1-x}$Ge$_x$ alloys.
One important limiting factor in the fabrication of strained Si
layers is the so-called {\em critical thickness}, $h_C$, the
thickness above which the strained layer relaxes by forming
topological defects, typically misfit dislocations  which originate
at the interface and propagate into the bulk.\cite{Wang}  Theoretical
estimates through an energy minimization criterion\cite{strain} 
show a sharp decay of $h_C$ with the alloy Ge content, from over
1000 \AA~ for $x=0.05$ to 100 \AA~ for $x=0.35$.  Recent advances in
epitaxial growth of Si/Ge heterostructures have led to the
fabrication of samples of 100 \AA~ thick Si layers over relaxed
Si$_{1-x}$Ge$_x$ alloys with $x$ up to 0.3,\cite{mobility} with
interfaces of extremely good quality and free of dislocations. 
Also, 150 \AA~ thick layers over $x=0.2$ alloys were recently
produced.\cite{strain}  In both cases, the layer thickness is below
(but very close to) the theoretical upper bound for $h_C$.

Substrate layers with relatively high Ge-content ($x \sim 0.2$) serve
two useful purposes in the context of QC architecture: (i) the
strained Si layer leads to improved behavior of exchange coupling
between donor impurity states; (ii) the relaxed alloy layer serves as
a barrier separating the Si host containing $^{31}$P donors from the
conducting  gates.\cite{Kane00}  Such values of alloy composition
$x$, leading to valley strain $\chi\sim -20$,  also impose some
limitations in the implementation of the Si-based QC.  The magnitude
of strain means an upper bound of around 200 \AA~ for the
thickness of the strained Si layer, a reduction in the hyperfine
coupling, and a reduction in the energy splitting between the ground
and the first excited state by a factor of 3 with respect to the
relaxed Si architectures.  Alloys with lower Ge content ($x \sim
0.05$) with valley strain parameter $\chi \sim -5$ provide comparable
electronic performance (actually somewhat improved performance in
terms of the strength of the hyperfine coupling) and greatly enhanced
values of $h_C$, over 1000 \AA.  This would obviously facilitate most
aspects of the design and fabrication of the Si layers.  On the
other hand, the low-Ge-content alloys do not provide a sufficiently
high potential barrier, thus one would have to use the more
conventional SiO$_2$ barriers between the strained Si and the top
gates.

Our results refer to structures of strained Si commensurately 
grown over relaxed random alloys.  The inevitable substrate alloy
disorder can pose some potential difficulties.  Indeed, some
randomness in the lattice spacing and thus in strain is bound to
occur at the alloy/Si interface.  However, this effect quickly decays
inside the Si layer. Since we consider donors located as far as 100
\AA~ ($\sim$ twenty lattice spacings in Si) away from the interface,
any alloy disorder  effect should be negligible, leading to a
uniformly strained Si host  for all practical purposes.  
On the other hand, if a device is based on an heterostructure 
predominantly formed by Si$_{1-x}$Ge$_x$/Si$_{1-y}$Ge$_y$ strained
layers (alloy on alloy), as considered in Ref.~\onlinecite{Vrijen},
and the donors are located inside an alloy layer (see
Fig.~\ref{fig:computers}(b)), random strain and local chemical
disorder would presumably play a much more important role in
determining the donor electronic properties, including exchange and
hyperfine couplings.  The electronic properties of a single impurity
is sensitive to local composition fluctuations, which will certainly
influence the performance  and reproducibility in the fabrication of
such devices. Another suggested alloy,\cite{Kane00} Si$_{1-x}$C$_x$,
presents a much larger lattice mismatch within the alloy layer,
leading to a much stronger randomness in local strain, which implies
a higher probability  of defects and smaller critical thickness if
one attempts to grow strained Si on top of such a substrate.  

We have demonstrated that it is possible to
achieve hydrogenic behavior for substitutional donors positioned
exactly in the same (001) lattice plane parallel to the interface
with the alloy substrate.  For the recently proposed\cite{Clark}
``bottom-up'' approach for Si QC fabrication, in which MBE growth is
followed by positioning individual P impurity on the surface with
the help of an STM tip, our result implies that small displacements
of the P atoms due to surface diffusion or hops among the Si atoms
forming dimers at the Si(001)$2\times1$ surface should be acceptable
for the reliability of  QC operations mediated by exchange
coupling.  The final stage of the ``bottom-up'' approach involves Si
overgrowth, encapsulating the deposited P donors.  This is
a crucial step for either relaxed or strained Si, since any
atomic-scale $z$-component added to the relative position ${\bf R}$
leads to oscillatory behavior and possibly strong reduction of $J$. 
In MBE growth, surface binding and elastic energy differences among
the various species lead to segregation effects, driving P atoms
from previously grown layers into newly deposited ones.  Segregation
may be reduced by lowering the growth temperature and by increasing 
the growth rate,\cite{Dehaese} but it may not be eliminated
entirely.  In a different, ``top-down'', approach, the P
array in Si is formed by ion-implantation techniques, which
implies a low degree of control over $R_z$ on a lattice-parameter
length scale.  Residual variations in the relative positions of the
donors must be corrected by individually calibrating the A and J
gates.  In other words, one must rely on probing and ``correcting''
each donor bound state through the surface gates.  In general, our
work establishes that Si QC design could tolerate some imprecision in
the positioning of the dopant atoms in a 2D layer parallel to the
interface, but any lack of precise control in the $z$ direction could
be fatal. 

We point out an additional advantage of strained over
relaxed Si according to our numerical results.  Following all
overgrowth stages, one A gate must be placed on top of each buried
impurity and one J gate in-between neighboring impurity pairs (see
Fig.~\ref{fig:computers}(a)).  We have shown that the exchange
coupling is enhanced for donor pairs  parallel to the interface in
negatively strained Si configurations. This allows for larger
inter-donor distances as compared to relaxed Si,  which is
particularly convenient in terms of fabricating and  accurately
positioning the external gates. 

Finally, we point out the obvious fact that, although we have
emphasized the relevance of our theory to Si QC architectures
throughout this paper, our work is of general validity
completely transcending QC considerations and context.  In
particular, we have developed a theory for calculating the donor
exchange energy in strained Si systems incorporating the quantum
interference effect arising from the multiple valley structure of Si
conduction band.  Our results should apply to any problem involving
Si donor exchange coupling considerations.

\acknowledgments
We thank Rogerio de Sousa for calling our attention to the approximate 
Heitler-London form for the exchange coupling we used previously. 
We also thank Bruce Kane, Ravin Bhatt, and Michelle Simmons for
useful conversations.  This work is supported by
ARDA and NSA.  BK acknowledges financial support from CNPq (Brazil).

\appendix
\section{Heitler-London Approximation for Donors in Si}

In this paper the Heitler-London approximation is employed to
calculate the exchange splitting $J$ between the ground singlet and
triplet states for two $^{31}$P donors embedded in strained Si.  Here
we derive the expression for $J$ in terms of the donor relative
position and the host band structure parameters.  Assuming the ground
donor electron wave functions to be $\psi({\bf r})$, the singlet and
triplet wave functions for donors located at the origin and ${\bf R}$
are  
\begin{eqnarray}
\psi_s ({\bf r}_1, {\bf r}_2) & = & \frac{1}{\sqrt{2(1 + S^2)}} 
\left[ \psi({\bf r}_1) \psi({\bf r}_2 - {\bf R})
+ \psi({\bf r}_1 - {\bf R}) \psi({\bf r}_2) \right] 
\times \frac{1}{\sqrt{2}} \left( |\uparrow \downarrow \rangle
- |\downarrow \uparrow \rangle \right) \,, \nonumber \\ 
\psi_t ({\bf r}_1, {\bf r}_2) & = & \frac{1}{\sqrt{2(1 - S^2)}} 
\left[ \psi({\bf r}_1) \psi({\bf r}_2 - {\bf R})
- \psi({\bf r}_1 - {\bf R}) \psi({\bf r}_2) \right] 
\times \frac{1}{\sqrt{2}} \left( |\uparrow \downarrow \rangle
+ |\downarrow \uparrow \rangle \right) \,, \nonumber 
\end{eqnarray}
where $S$ is the overlap integral
\begin{equation}
S = \int d^3{\bf r} \; \psi^*({\bf r}) \psi({\bf r} - {\bf R}) \,.
\label{eq:s}
\end{equation}
The Hamiltonian that governs this two-electron system is
\begin{equation}
H = \sum_{i=1}^2 \left\{ K_i - {e^2\over \epsilon} \left[
{1 \over |{\bf r}_i|} + {1 \over |{\bf r}_i-{\bf R}|} 
\right] \right\} + {e^2\over \epsilon} \left( 
{1 \over |{\bf r}_1-{\bf r}_2|} + {1 \over R} \right) \,.
\end{equation}
The energy splitting between the ground singlet and triplet states
within the Heitler-London approximation is\cite{Slater}
\begin{eqnarray}
J({\bf R}) & = & E_t - E_s = \langle \psi_T |H| \psi_T \rangle
- \langle \psi_S |H| \psi_S \rangle \nonumber \\
&= & \frac{2}{1 - S^4} (H_0 S^2 - H_1) \,. \nonumber
\end{eqnarray}
The terms appearing both here and in
(\ref{eq:hl}) are 
\begin{eqnarray}
H_0 & = & \int d^3{\bf r}_1 d^3{\bf r}_2 |\psi^* ({\bf r}_1)|^2 {\cal
O } |\psi^* ({\bf r}_2 - {\bf R})|^2 \,,
\label{eq:h0} \\
H_1 & = & \int d^3{\bf r}_1 d^3{\bf r}_2 \psi^* ({\bf r}_1)
\psi^* ({\bf r}_2 - {\bf R}) {\cal O} \psi ({\bf r}_1 - {\bf R}) \psi 
({\bf r}_2) \,,
\label{eq:h1}
\end{eqnarray}
with
\begin{equation}
{\cal O} = {e^2\over \epsilon} \left( -{1 \over |{\bf r}_1-{\bf R}|}
-{1 \over |{\bf r}_2|} 
+{1 \over |{\bf r}_1-{\bf r}_2|} +{1\over R} \right) \, ,
\label{eq:O}
\end {equation}

Replacing the ground state wave functions of the form (\ref{eq:sim})
for $\psi$ in the expressions above leads to sums over integrals
which involve different envelope functions and phases coming from
the Bloch wave functions.  For example, for the overlap (\ref{eq:s})
one gets 
\begin{eqnarray}
S & = & \sum_{\mu \nu \bf K \bf K'} A_{\mu\nu{\bf K}{\bf K'}} \int
d^3{\bf r}_1 F_\mu ( {\bf r}_1) F_\nu ({\bf r}_1- {\bf R}) e^{-i({\bf
k}_\mu+{\bf K} )  \cdot {\bf r}_1 } e^{i ({\bf k}_\nu+ {\bf K'})
\cdot({\bf r}_1 - {\bf R})}~, \nonumber \\
& = & \sum_{\mu \nu \bf K \bf K'} A_{\mu\nu{\bf K}{\bf K'}} \;
e^{-i {\bf k}_\nu \cdot {\bf R}} \int d^3{\bf r}_1 F_\mu ( {\bf
r}_1) F_\nu ({\bf r}_1- {\bf R}) e^{i ({\bf k}_\nu - {\bf k}_\mu
+ {\bf K'} - {\bf K}) \cdot {\bf r}_1} \,,
\end{eqnarray} 
where 
$$A_{\mu\nu{\bf K}{\bf K'}} = \alpha_{\mu}^* \alpha_{\nu} \,
c^{\mu \, *}_{\bf K} c^{\nu}_{\bf K'}\,.$$ 
Following the
approximations suggested in Ref.~\onlinecite{Andres}, we neglect all
integrals which contain rapidly oscillating phases {\em in the
integrated variables} ${\bf r}_1$ or ${\bf r}_2$. The remaining
integrals involve different anisotropic  envelope functions, and may
be cast into forms similar to those appearing in the H$_2$ molecule
HL problem through a coordinate rescaling along the three Cartesian
directions  by the respective effective Bohr radii. For the specific
example above, we get: 
\begin{equation}
S = \sum_{\mu {\bf K}} A_{\mu \mu{\bf K}{\bf K} } \; s_\mu
\, e^{-i  {\bf k}_\mu \cdot{\bf R}}
\end{equation}
with 
\begin{equation}
s_{\mu}=e^{-{\cal R}_\mu}(1+{\cal R}_\mu+{\cal R}_\mu^2/3).
\label{eq:over} 
\end{equation}
Equation (\ref{eq:over}) is exactly the hydrogenic expression for the
overlap, but evaluated at the  rescaled internuclear distance which,
for $\mu=z$, is $\bbox{\cal R}_z = (R_x/a,R_y/a,R_z/b)$. Integrals
appearing in (\ref{eq:h0}) and (\ref{eq:h1}) contain a
distance-dependent  denominator which, at large internuclear
separations, may be replaced by the value appropriate for the line
joining the two donors.\cite{Andres}  For these terms it is
convenient to define 
\begin{equation} 
f(\theta_{\mu}) = {e^2 \over{\epsilon a}} \left[ {a^2\over
b^2} \cos^2 \theta_{\mu} +\sin^2 \theta_{\mu} \right]^{1/2}
\label{eq:ang}
\end{equation}
where $\theta_{\mu}$ is the polar angle between the $\mu$-direction
and ${\bf R}$. 

Taking all these considerations into account, we arrive at the final
expression for the exchange coupling $J({\bf R})$:
\begin{equation} 
J({\bf R}) = \sum_{\mu, \nu} \left[ \sum_{{\bf K},{\bf K'}}
|c^\mu_{\bf K}|^2  |c^\nu_{\bf K'}|^2 e^{i({\bf K} - {\bf K'})\cdot
{\bf R}} \right] |\alpha_\mu|^2 |\alpha_\nu|^2 {\cal J}_{\mu \nu}
({\bf R}) \cos ({\bf k}_{\mu}-{\bf k}_{\nu})\cdot {\bf R}\,.
\label{eq:exch}
\end{equation}
The expression for the kernel $\cal J$ is
\begin{equation}
{\cal J}_{\mu \nu}= {2\over{1-S^4}} [s_\mu s_\nu (2C_1 + C_2) -
(s_\mu  v_\nu + v_\mu s_\nu + j_{\mu \nu})]
\label{eq:all}
\end{equation}
where
\begin{eqnarray}
C_1 & = & \sum_{\lambda}|\alpha_\lambda|^2 c_\lambda \,, \\
\label{eq:c1}
C_2 & = & \sum_{\lambda\gamma}|\alpha_\lambda|^2 |\alpha_\gamma|^2
c'_{\lambda\gamma} \,, 
\label{eq:c2}
\end{eqnarray}
$c_\lambda = f(\theta_\lambda) [-1/{\cal R}_\lambda +
e^{-2{\cal R}_\lambda} (1+1/{\cal R}_\lambda)]$, 
$c'_{\lambda \lambda} = f(\theta_\lambda) [1/{\cal R}_\lambda - 
e^{-2{\cal R}_\lambda}(1/{\cal R}_\lambda
+11/8+(3/4) {\cal R}_\lambda + (1/6) {\cal R}_\lambda ^2)]$,~
$v_\mu = f(\theta_\mu) [- e^{-{\cal R}_\mu}(1+{\cal R}_\mu)]$,~ 
$s_\mu$ is given in (\ref{eq:over}), 
and 
$j_{\mu \mu} = f(\theta_\mu)j_H({\cal R}_\mu)$, with $j_H(r)$ as given
in (B5) of Ref.~\onlinecite{Andres}.  As in Ref.~\onlinecite{KHD}, we
make the following additional assumption regarding the off-diagonal
elements of $c'$ and $j$: $j_{\mu \nu} = \sqrt{j_{\mu \mu}
j_{\nu\nu}}$ and $c'_{\lambda\gamma} =
\sqrt{c'_{\lambda\lambda}c'_{\gamma\gamma}}$. 

\begin{figure}
\centerline{
\epsfxsize=3.6in
\epsfbox{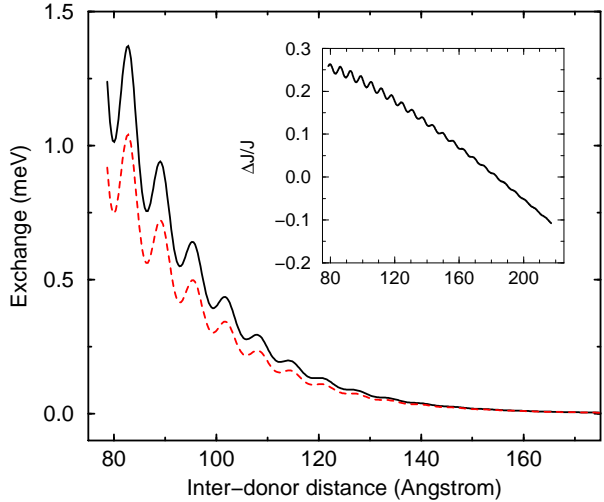}}
\vspace*{0.1in}
\protect\caption[Comparison of Heitler-London and Andres results
for exchange]
{\sloppy{
Comparison of the results from calculations with the
Heitler-London approximation (solid curve) and its approximate form
(dashed curve) for the donor exchange in bulk relaxed Si.  Without
loss of generality for the comparison, the two donors here are placed
along [100] axis. The inset gives the difference
between these two approaches in units of the HL value for exchange as
a function of the inter-donor distance.  At about 180 \AA~ donor
separation there is a sign change in the exchange difference $\Delta
J$, while in the range of distances of practical interest, from 100
to 200 \AA, the relative difference remains small.  
}}
\label{fig:HL-A}
\end{figure}

The expression (\ref{eq:exch}) for exchange $J({\bf R})$ is formally
equivalent to the $J({\bf R})$ obtained by Andres {\it et
al}~ \cite{Andres} under  the additional approximation $\phi_\mu(\bf
r) \sim e^{i {\bf k}_{\mu}\cdot {\bf r}}$, thus $c^\mu_{\bf K} =
\delta_{{\bf K},0}$, but with ${\cal J}_{\mu \nu}$ replaced by  
\begin{equation}
j_{\mu \nu} ({\bf R}) = \int d^3{\bf r}_1 d^3{\bf r}_2 F_{\mu}^* ({\bf
r}_1) F_{\nu}^* ({\bf r}_2 - {\bf R}) \frac{e^2}{\epsilon |{\bf r}_1 -
{\bf r}_2|} F_{\mu} ({\bf r}_1 - {\bf R}) F_{\nu} ({\bf r}_2) \,.
\label{eq:approx}
\end{equation}
This result emerges from an approximation for the exchange 
coupling commonly adopted in the
literature,\cite{Privman,KHD,Andres} in which only the exchange
integral of the Coulomb electron-electron  repulsion term is
included.  To our knowledge, no formal justification for this
assignment is available for electrons bound to donor impurities via
Coulomb interaction.  However, results presented in
Fig.~\ref{fig:HL-A} show that this is actually an acceptable
approximation if the inter-donor separation is between 100 and 200
\AA~ in Si, where relative changes $|\Delta J({\bf R})|/J({\bf R})$
are at most 20\%, keeping the  same qualitative features and trends
reported here for all strain  and $\bf R$ values investigated.
In fact, it is possible that the approximate expression
obtained using (\ref{eq:approx}), which is positive-definite, 
instead of (\ref{eq:all}), becomes more reliable
than HL in the $R\to \infty$ limit, given the artifact of the HL
expression for the exchange splitting in H$_2$ molecule that
predicts an unphysical negative $J$ for $R$ beyond 50 Bohr
radii.\cite{HF,angelescu}

%

%

%

%

%

%

\end{document}